\documentclass[twocolumn,twoside,slac_two]{revtex4}
\usepackage{graphicx}
\usepackage{fancyhdr}
\usepackage{epstopdf}
\newcommand{\met}{\,/\!\!\!\!E_{T}}

\newcommand{\sht}{H_T}

\newcommand{\ttbar}{{t\bar{t}}}
\newcommand{\tprime}{t^\prime}
\newcommand{\tprimebar}{\bar{t}^\prime}
\newcommand{\bprime}{b^\prime}

\begin{document}

\title{Search for a heavy top $\tprime \Rightarrow Wq$ in top events}

%

\author{D. Cox, on behalf of the CDF Collaboration}
\affiliation{Department of Physics, University of California-Davis, Davis, CA 95616, USA}

\begin{abstract}
We present a search for a massive quark ($\tprime$) decaying to $Wq$ and thus mimicking the top quark decay signature in data collected by the CDF II detector corresponding to 2.8 fb$^{-1}$. We use the reconstructed mass of the $\tprime$ quark and the scalar sum of the transverse energies in the event to discriminate possible new physics from Standard Model processes, and set limits on a standard 4th generation $\tprime$ quark.
\end{abstract}

\maketitle

\thispagestyle{fancy}


\section{Introduction}
In this study we investigate whether present data from the CDF detector allow or preclude the production of a hypothetical new quark which decays to a final state with a high-$p_{T}$ lepton, large $\met$, and multiple hadronic jets having large total transverse energy $E_{T}$ thus mimicking top quark pair event signatures.
We refer to the hypothetical new quark as $\tprime$ for brevity, although such a signature could be a standard fourth-generation up-type heavy quark or any up-type quark.
A number of theoretical models advocate a fourth chiral generation of massive fermions with the same quantum numbers as ordinary ones \cite{democratic,gut,little-higgs,n=2,bm}.
Precision measurements from LEP exclude a light fourth neutrino $\nu_4$ with mass $m(\nu_4) < m_Z/2$, where $m_Z$ is the mass of the $Z$ boson. A fourth generation neutrino cannot be too heavy due to sizeable radiative corections. Despite this, reasonable constructions of a fourth generation exist that remain viable \cite{constructions}.
In many theoretical models the addition of a fourth generation relaxes present bounds on the Higgs. In addition frequently a small mass splitting between new heavy quarks $\tprime$ and $\bprime$ is preferred, such that $m(\bprime) + m(W) > m(\tprime)$ and $\tprime$ decays predominantly to $Wq$ (a $W$ boson and a down-type quark $q = d,s,b$) \cite{frampton}.
For the purposes of this analysis we assume that the new quark is pair-produced strongly, has mass greater than the top quark, and decays promptly to $Wq$ final states. The data we use in the analysis corresponds to 2.8 fb$^{-1}$ of integrated luminosity collected at the Fermilab Tevatron in $p\bar{p}$ collisions at $\sqrt{s} = 1.96$ TeV.

\section{Event Selection}
The CDF II detector is described in detail in Ref. \cite{cdf_detector}. We can parameterize coordinates in the detector using the azimuthal angle $\phi$ and the pseudorapidity $\eta = -\ln[\tan(\theta/2)]$, where $\theta$ is the polar angle measured from the proton beam direction. The transverse energy $E_T$ is defined as  $E\sin\theta$, where $E$ is the energy deposited in a calorimeter cluster. The transverse momentum $p_T$ of a track is the component of the track momentum transverse to the beam-line. The missing transeverse energy $\met$ is the magnitude of the vector defined as -$\Sigma_iE^{i}_T\hat{n}^{i}_{T}$, where $\hat{n}^{i}_{T}$ is the transverse component of the unit vector pointing from the interaction point to the calorimeter tower $i$. This is corected for the $p_T$ of muons, which do not deposit all of their energy in the calorimeter, and tracks which point to uninstrumented regions in the calorimeter. 

Highly energetic quarks (such as the down-type quark produced in the hypothesized $\tprime$ decay) undergo fragmentation that results in jets of hadronic particles. Jet candidates are reconstructed using the calorimeter towers with corrections to improve the accuracy of the energy estimation and are required to have $\met > 15$ GeV and $|{\eta}| < 2.5$.

Candidate events for this $\tprime$ search are required to have exactly one isolated electron or muon with $p_T \geq 20$ GeV/c, $\met \geq 35$ GeV, at least four jets with $E_T \geq 20$ GeV.

To reduce the contribution of the QCD background we also require a lead jet with $E_T \geq 60$ GeV and two cuts one in the $\Delta\phi$ between the corrected $\met$ and the lepton vs $\met$ plane (requiring $\Delta\phi \geq A_1 - (1/B_{1})\met$ where $A_1$ = 4.408; $B_1$ = 6.11) and one in the $\Delta\phi$ between $\met$ and lead jet vs $\met$ plane (requiring $\Delta\phi \geq A_2 - (1/B_2)\met$ where  $A_2$ = 1.888; $B_2$ = 21.6). 
The dominant contributing backgrounds after this event selection are from electroweak process as well as $\ttbar$ pair production. Electroweak processes are dominated by $W$ + jets. For the $\ttbar$ background  we use Monte Carlo (MC) generated using an assumed top quark mass of 175 GeV. The QCD background is modeled using a sample of data where the lepton ID cuts have been reversed. Other backgrounds (including $Z$+jets, $WW$+jets, $WZ$+jets and single top events) have a smaller rate than $W$ + jets and in addition have been found to have similar kinematic distributions to $W$ + jes and so are modeled as one background using the $W$ + jets model.

\section{Search Technique}
We utilize the fact that in our regime of interest the $\tprime$ decay chain is identical to that of the top quark to reconstruct the $\tprime$ mass in the same way as is done in the top quark mass measurement analyses. We use the template method for top quark mass reconstruction \cite{template} based on the best $\chi^{2}$-fit to the kinematic properties of the final top (or $\tprime$) decay products. 
The $\chi^2$ is given by the following expression: 

\begin{eqnarray}
\chi^2 & = & \sum_{i=\ell,4 jets} \frac{(p_T^{i,fit} - p_T^{i,meas})^2}{\sigma^2_i} \nonumber \\
& & + \sum_{j=x,y} \frac{(p_j^{UE,fit} - p_j^{UE,meas})^2}{\sigma^2_j} \nonumber \\
& & + \frac{(m_{jj}-m_W)^2}{\Gamma_W^2} + \frac{(m_{\ell\nu}-m_W)^2}{\Gamma_W^2} \nonumber \\
& & + \frac{(m_{bjj}-m_t)^2}{\Gamma_t^2} + \frac{(m_{b\ell\nu}-m_t)^2}{\Gamma_t^2} \nonumber \\\label{Chi2}
\end{eqnarray}
where the invariant masses of the $W$ decay products $m_{jj}$ and $m_{\ell\nu}$ are constrained to the pole mass of the $W$ boson, and the masses of top and anti-top ($\tprime$ and $\bar{\tprime}$) quarks are required to be same. The jet and lepton energies as well as the unclustered energy ($UE$) are allowed to float within their resolution uncertainties. The transverse component of the neutrino momentum is determined as the negative sum of the lepton, jet and unclustered transverse
energies:
 \begin{equation}
  \vec{p}_T^{ \nu} = - (  \vec{p}_T^{ \ell}  + \sum \vec{p}_T^{ jet}  + \vec{p}_T^{ UE} ).
\label{neu}
\end{equation}

For each event there are total 4!/2 = 12 combinations of assigning 4 jets to partons. In addition, there are two solutions to account for the unknown z-component of the neutrino momentum. After minimization of the $\chi^2$ expression, the combination with the lowest $\chi^2$ is selected and the value of $m_t$ is declared to be the reconstructed mass $M_{reco}$ of top (or $\tprime$).

We use the observed distributions of the $M_{reco}$ and total transverse 
energy in the event 
\begin{equation}
\sht = \sum_{jets} E_T + E_{T,\ell} + \met \nonumber
\label{Ht}
\end{equation}
to distinguish the $\tprime$ signal from the backgrounds by fitting it to a combination of $\tprime$ signal,
top, electroweak background, and QCD background shapes.

We use a binned in $\sht$ and $M_{reco}$ likelihood fit to extract the $\tprime$ signal and/or set an upper limit on its production rate. We chose to use bins of 25 GeV in both $H_T$ and $M_reco$ with $H_T$ in 26 bins from 150 to 800 GeV and $M_reco$ in 16 bins from 100 to 500 GeV.
The likelihood is defined as the product of the Poisson probabilities for observing $n_i$ events in 2D bin $i$ of ($\sht,M_{reco}$):
\begin{eqnarray}
    {\cal L}(\sigma_{\tprime}|n_i) = \prod_i P(n_i|\mu_i) \ \ \  \nonumber
\end{eqnarray}

The expected number of events in each bin, $\mu_i$, is given by the sum over all sources, indexed 
by $j$
\begin{eqnarray}
     \mu_i = \sum_j L_j \sigma_j \epsilon_{ij}  \ \ \  \nonumber
\label{mu}
\end{eqnarray}
Here $L_j$ is the integrated luminosity, $\sigma_j$ is the cross section, and $\epsilon_{ij}$ is the efficiency per bin of ($\sht,M_{reco}$).

We calculate the likelihood as a function of the $\tprime$ cross section, and use Bayes' Theorem to convert it into a posterior density in $\sigma_{\tprime}$.  We can then use this posterior density to set an upper limit on  (or if we get lucky, measure) the production rate of $\tprime$.

The production rate for $W$+jets is a free parameter in the fit. Other parameters are related to systematic errors and treated in the likelihood as nuisance parameters constrained within their expected (normal) distributions. 

We adopt the profiling method~\cite{prev_Gen4} for dealing with these parameters, 
i.e. the likelihood is maximized with respect to the nuisance parameters. Taking this into account the likelihood takes the following expression:  
\begin{eqnarray}
    {\cal L}(\sigma_{\tprime}|n_i) = \prod_{i,k} P(n_i|\mu_i) 
           \times  G(\nu_k |\tilde{\nu}_k,\sigma_{\nu_k}) \ \ \  \nonumber
\label{systlikelihood}
\end{eqnarray}
where $\nu_k$ are the nuisance parameters, such as $\sigma_{\ttbar}$, $L_j$ and etc. ${\tilde{\nu}}_k$ are their central nominal values and $\sigma_{\nu_k}$ are their uncertainties. $G$ is a Gaussian function centered at ${\tilde{\nu}}_k$  of width $\sigma_{\nu_k}$.

\section{Systematic Errors}
The sensitivity to $\tprime$ depends on knowing accurately the
distribution of ($\sht,M_{reco}$) in data. We consider several sources of uncertainty including the jet energy scale, $W$+jets $Q^2$ Scale, initial and final state radiation (ISR and FSR), Parton Distrubiton Function (PDF) uncertaintity and others.

\subsection{Jet Energy Scale}
The sensitivity to $\tprime$ depends on knowing accurately the
distribution of ($\sht,M_{reco}$) in data. One of the largest sources of 
uncertainty comes from a factor that has a large effect 
on the shape of the kinematic distribution, the jet energy scale.  
Jets in the data and MC are
corrected for various effects as described in~\cite{jes}, 
leaving some residual uncertainty.

This uncertainty results in possible shifts in the $\sht$ and 
$M_{reco}$ distributions for both new physics and standard 
model templates. 
We take this effect into account by generating templates 
with energies of all jets shifted upwards by one standard 
deviation (+1 templates) and downwards (-1 templates) respectively.

We then use a template morphing technique that was developed in 2005 for a previous version of this analysis.
We interpolate and extrapolate the
expectation value $\mu_i$ at each bin $i$ as follows:
\begin{equation}
\mu_i = \mu_{0,i} + \nu_{JES} \cdot(\mu_{+1,i} - \mu_{-1,i}) / 2 \nonumber
\label{mu_JES}
\end{equation}
where $\mu_{0,i}$ is the nominal expectation value, $\mu_{-1,i}$ and 
 $\mu_{+1,i}$ are the expectation values from (-1) and (+1) templates
respectively,and$\nu_{JES}$ is the nuisance parameter representing the relative shift in jet energy scale:
\begin{equation}
  \nu_{JES} = \frac{\Delta_{JES}}{\sigma_{JES}} \nonumber
\label{nuJES}
\end{equation}

It enters the likelihood~(\ref{systlikelihood}) 
as a Gaussian constraint penalty term: $G(\nu_{JES}|0,1) = \frac{1}{\sqrt{2\pi}} e^{-\nu_{JES}^2/2}$.
\subsection{$W$+jets $Q^2$ Scale}
The effect of the choice of the appropriate $Q^2$ scale for $W$+jets production is evaluated by measuring the resulting change in the measured $\tprime$ cross section given that $\tprime$ exists. The $Q^2$ scale is varied to $2Q^2_{nominal}$ and $\frac{1}{2}Q^2_{nominal}$ the expected change in the measured cross section is then interpreted as the uncertainty on the $\tprime$ cross section itself.  We measure this shift as a function of the $\tprime$ cross-section by drawing pseudoexperiments from shifted templates and fitting them to the nominal distribution. The resulting shift is fitted to a linear function of that $\tprime$ cross-section and is incorporated into the likelihood as an additive parameter to the $\tprime$ cross section, so that the $\tprime$ contribution to the expectation value $\mu_{i}$~(\ref{mu}) in bin $i$ becomes
\begin{equation}
 \mu_{i,\tprime} = L_{\tprime} ( \sigma_{\tprime} + \nu_{Q^2} ) \epsilon_{i,\tprime} \nonumber
\end{equation}
where $\nu_{Q^2}$ is constrained by a gaussian with a width, that is a half of the largest of the upwards or downwards shifts for each mass of the $\tprime$. 

The $Q^2$ systematic uncertainty for the different $\tprime$ masses are shown in Table~\ref{tab:q2}.
\begin{table}[h]
    \begin{tabular}{c|c} \hline
         $m(\tprime)$  & Systematic Uncertaintity \\
         (GeV)         & (pb) \\ \hline
   180   &     0.065 \\
   200   &     0.044 \\
   220   &     0.021 \\
   240   &     0.011 \\
   260   &     0.013 \\
   280   &     0.009 \\
   300   &     0.007 \\
   320   &     0.005 \\
   340   &     0.006 \\
   360   &     0.005 \\
   380   &     0.003 \\
   400   &     0.003 \\
   450   &     0.003 \\
   500   &     0.002 \\ \hline
    \end{tabular}
  \caption{ $Q^2$ Systematic error for each $\tprime$ mass point.}
  \label{tab:q2}
\end{table}
\subsection{ISR and FSR}
We varied the amount of initial- and final-state radiation together, i.e. shifting both up or both down. We generated samples with more ISR and FSR as well as some with less ISR and FSR. We refer to these samples as IFSR more and IFSR less. We generated samples for $\tprime$ with masses of 250, 300 and 350 GeV which brackets the region where we expect to be able to place our exclusion limit.

The resulting effect is treated in a similar way to the $Q^2$ systematic. Templates are made for each of these mass points. Pseudoexperiments are thrown with the shifted top and $\tprime$ IFSR samples, where the shift is set to be the same for top and $\tprime$. We then fit the obtained cross-section shift using a linear function of the $\tprime$ cross-section.

The resulting shifts are shown in Table~\ref{tab:ifsr}. We add the resulting shifts in quadrature with the $Q^2$ error in  the likelihood.
\begin{table}[h]
    \begin{tabular}{c|cc} \hline
         $m(\tprime)$  & \multicolumn{2}{c}{IFSR} \\
         (GeV)         & offset & slope \\ \hline
   180   &     0.125 & 0.026 \\
   200   &     0.125 & 0.024 \\
   220   &     0.125 & 0.022 \\
   240   &     0.110 & 0.020 \\
   260   &     0.080 & 0.018 \\
   280   &     0.060 & 0.017 \\
   300   &     0.035 & 0.014 \\
   320   &     0.025 & 0.011 \\
   340   &     0.015 & 0.009 \\
   360   &     0.010 & 0.008 \\
   380   &     0.007 & 0.007 \\
   400   &     0.005 & 0.006 \\
   450   &     0.004 & 0.005 \\
   500   &     0.003 & 0.004 \\
    \end{tabular}
  \caption{ Columns 2+3: Shift (in picobarns) in apparent $\tprime$ cross section due to a shift in the initial- and final-state radiation up or down}
  \label{tab:ifsr}
\end{table}
\subsection{QCD Background}
The QCD background shape is modeled from a sample of data in which the electron cuts have been reversed. The QCD normalization is obtained by fitting the background (electroweak, top, and QCD) distrubtions to the data with the missing $E_T$ cut removed and then computing how much remains after all cuts are applied. As most of the QCD is expected to be found at low missing $E_T$.

We investigated using leptons that fail the isolation requirements to model our QCD background. This gave an excellent description of the jet $E_T$ spectra but a very poor description of the lepton $p_T$ distributions, which were much steeper in this model than expected from the data.
On the other hand the sample requiring the opposite of the electron cuts seems to give a reasonable description of most of the data kinematic distributions but has only very limited statistics; this means that the QCD templates need to be scaled up.
Cutting very hard on the leading jet $E_T$ removes most of the QCD background which makes our fit rather insensitive to the QCD modeling.
The relative normalization uncertainty is taken to be 50\%, due to our lack of confidence in our model and normalization method.  With our QCD veto cuts it turns out to change the fit by a negligible amount whether we constrain QCD or let it float.
The uncertainty is represented by a Gaussian-constrained
parameter in the likelihood.  The QCD background has a negligible 
effect on the $\tprime$ limit.
\subsection{Integrated Luminosity}
The integrated luminosity uncertainty is taken to be 5.9\%, and is represented by an additional gaussian-constrained parameter multiplying all contributions except for the QCD background, which is normalized from data.
\subsection{Lepton ID}
We have two components for lepton ID. First is the efficiencies for the individual electrons and muons. We multiply each lepton type by the associated efficiency and gaussian constrain it within the error on the efficiency.

Second is the uncertainty on the lepton ID efficiency data/MC scale factor, which is of 2\%, and taken as correlated across lepton types since it is due to the presences of multiple jets in an event. We add it in quadrature with the luminosity error, which is also correlated across lepton types, and include it with a gaussian constraint into the likelihood.
\subsection{PDF Uncertainty}
The PDFs are not precisely known, and this uncertainty leads to a corresponding uncertainty in the predicted cross sections, as well as the acceptance.

This effect is evaluated on both the top and the t' MC samples. The method consists in re-weighting the existing MC samples by the relative PDF weights given the parton momentum fractions ($x_1$, $x_2$) and $Q^2$ of the generated interaction.

46 eigenvectors are considered. We look at the difference between pairs of the CTEQ6M PDFs and add up these in quadrature. We then consider the difference between the two MRST72 and CTEQ5L PDF sets. If this is smaller than the 20 PDF sets uncertainty, we drop it. If it is larger, we add it in quadrature. To investigate the effect of $\alpha_s$ we look at the difference between the MRST72 and MRST75 PDF sets and add this in quadrature to the above errors.

The final PDF uncertaintes are given for each $\tprime$ mass point as well as for top in Table~\ref{tab:pdf}. A common conservative systematic error is added in quadrature to all other multiplicative factors and it taken as 1.1$\%$ for all templates.
\begin{table}[h]
  \begin{tabular}{c|cc} \hline
\multicolumn{3}{c}{top} \\
175 & +0.0110 &  -0.0112 \\ \hline
\multicolumn{3}{c}{tprime} \\
180 & +0.007 & -0.008 \\
200 & +0.004 & -0.005 \\
220 & +0.005 & -0.005 \\
240 & +0.003 & -0.003 \\
260 & +0.003 & -0.003 \\
280 & +0.002 & -0.003 \\
300 & +0.001 & -0.003 \\
320 & +0.001 & -0.002 \\
340 & +0.002 & -0.002 \\
360 & +0.003 & -0.002 \\
380 & +0.002 & -0.002 \\
400 & +0.005 & -0.002 \\
450 & +0.004 & -0.005 \\
500 & +0.015 & -0.013 \\ \hline
  \end{tabular}
  \caption{PDF uncertainty on top and $\tprime$ calculated by reweighting the events according to the probability (given the various PDFs) of finding an up and down quark with appropriate momentum fractions.}
  \label{tab:pdf}
\end{table}
\subsection{Theory Uncertainty}
The theory uncertainty in the $\tprime$ cross section is about 10\% 
(see Table~\ref{tab:theory_curve}), which is mainly 
due to uncertainty in PDFs ($\sim 7\%$). The other effect
comes from 
uncertainty in the choice of the $Q^2$ scale~\cite{Cacciari:2003fi}.

We take the theory uncertainty in $\ttbar$ cross section 
fully correlated with the one of $\tprime\tprimebar$, and introduce 
it into the likelihood as a single nuisance parameter: 
$\nu_{theory} = \nu_{theory}(m_\tprime)$.
\section{Results and Conclusion}
\begin{figure}[h]
\centering
\includegraphics[width=80mm]{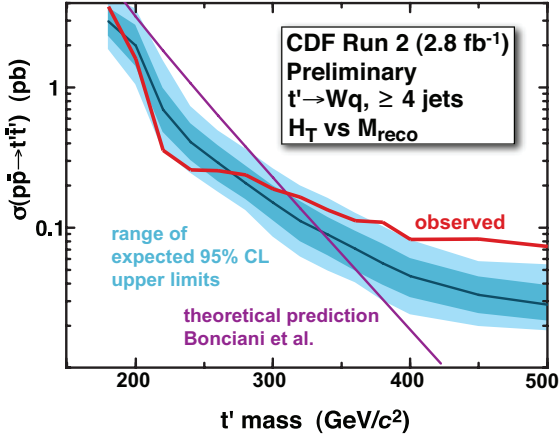}
  \caption{Upper limit, at 95\% CL, on the production rate for $\tprime$
           as a function of $\tprime$ mass (red).
           The purple curve is a theoretical 
           cross section. The dark blue band is the range of 
           expected 95\% CL upper limits within one standard deviation. The light blue band represents two standard deviations}
  \label{fig:result}
\end{figure}

We tested the sensitivity of our method by drawing pseudoexperiments from standard model distributions, i.e. assuming no $\tprime$ contribution. The ranges of the expected 95\% CL upper limits with one and two standard deviation bandwidth are shown in Figure~\ref{fig:result}.
The purple curve is the theoretical prediction~\cite{Bonciani:1998vc,Cacciari:2003fi}, the values of which are given in Table~\ref{tab:theory_curve}. The lower $\sigma_{min}$ and upper $\sigma_{max}$ limits are obtained using the CTEQ6M family of parton density functions with uncertainties, together with the study of the scale uncertainty~\cite{MMangano}.
\begin{table}[ht]
  \begin{tabular}{c|ccc} \hline
  $m(\tprime)$  (GeV)  & $\sigma_{min}$ (pb)  & $\sigma_{center}$ (pb)   & $\sigma_{max}$ (pb) \\ \hline

  180.0     &     4.9938    &    5.7476    &    6.2396   \\
  200.0     &     2.7815    &    3.1898    &    3.4525   \\
  220.0     &     1.5926    &    1.8236     &   1.9710   \\
  240.0      &    0.9299    &    1.0647     &   1.1515   \\
  260.0     &     0.5499    &    0.6302     &   0.6828   \\
  280.0     &     0.3281    &    0.3769    &    0.4096   \\
  300.0     &     0.1968    &    0.2268     &   0.2475   \\
  320.0    &      0.1183    &    0.1370     &   0.1502   \\
  340.0    &      0.0711    &    0.0828     &   0.0914   \\
  360.0     &     0.0426    &    0.0500     &   0.0555   \\
  380.0    &      0.0255    &    0.0301     &   0.0337   \\
  400.0    &      0.0152    &    0.0181     &   0.0204   \\ \hline
  \end{tabular}
  \caption{Theory values of $\tprime$ cross section for given mass~\cite{Cacciari:2003fi,Bonciani:1998vc}.}
  \label{tab:theory_curve}
\end{table}
From Figure~\ref{fig:result} it follows that given no $\tprime$ presence, this method is on average sensitive to setting an upper limit at 311 GeV $\tprime$ mass. The red curve in Figure~\ref{fig:result} shows the final result expressed as a 95\% CL upper limit on the $\tprime$ production rate as a function of $\tprime$ mass. Table~\ref{tab:result} shows the individual calculated limits along with expected limits from pseudo-experiments for reference.

Based on these results we exclude at 95\% CL a $\tprime$ quark with a mass below 311 GeV, given that the true top mass is 175 GeV. Of course, our measurement of the top mass may have been affected by the presence of a higher mass $\tprime$ and thus we should treate these conclucsions with care.

\begin{figure*}[t]
\centering
\includegraphics[width=65mm]{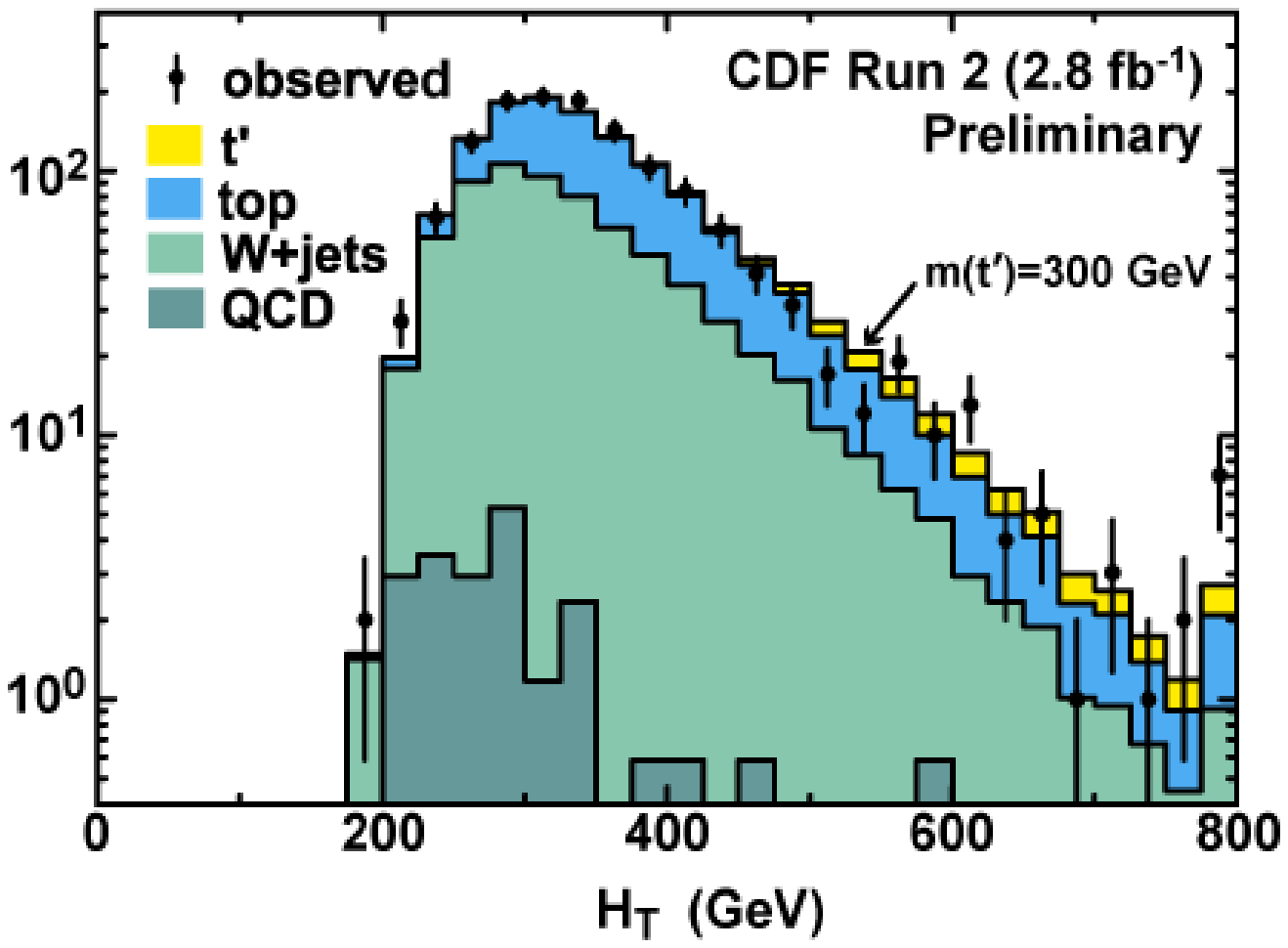}
\includegraphics[width=65mm]{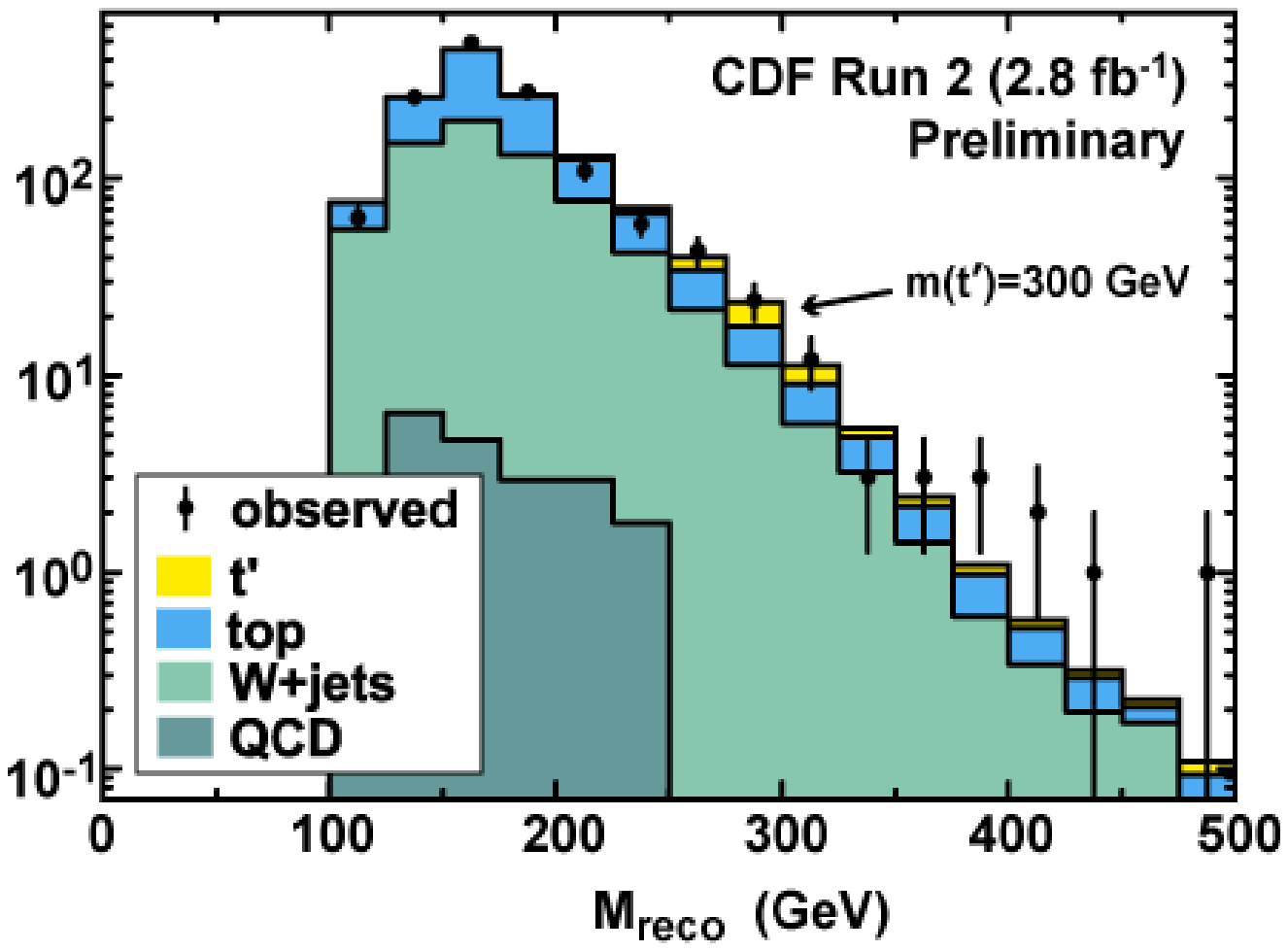}
\caption{$\sht$ (left \& $M_{reco}$ (right) distriubtions showing the results of the fit for m($\tprime$) = 300 GeV. The normalizations of the various sources and distortions of the kinematic distributions due to systematic effects are those corresponding to the maximum likelihood when the cross section for t' is set to its 95\% CL upper limit.}
\label{fig:htmt300}
\end{figure*}

\begin{figure}[h]
\centering
\includegraphics[width=60mm]{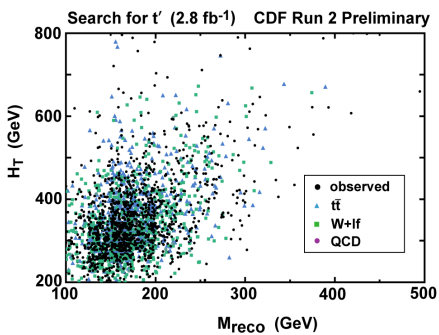}
\caption{2D distriubtion of $\sht$ vs $M_{reco}$ distribution showing hte data (black points) and the fitted number of background events: QCD (dark cyan circles), $W$ + jets (green squares), and $\ttbar$ (blue triangles).}
\label{fig:2d}
\end{figure}

The 2D-distriubtion of ($\sht$, $M_{reco}$) is shown in Figure~\ref{fig:2d}. Distributions of $\sht$ and $M_{reco}$ showing the result of the fit for $m(\tprime)$=300 GeV are shown in Figure~\ref{fig:htmt300}

To determine if the data show any evidence of an excess in the tails of $\sht$ and $M_{reco}$, we decided \textit{a priori} to count the number of events in groups of $n \times n$ of our standard 25 GeV bins in these quantties, and compare with the number predicted from a zero-signal fit to the full two dimensional spectrum. For each $n \times n$ bin one can then calculate the p-value for having observed that number or greater, given the prediction. If a significant effect is observed, one can calculate an overall p-value which is the probability that one would observe a p-value at least as significant as the most significant $n \times n$ bin or greater; this takes into account both the trials factor and the effect of systematic errors.
Table~\ref{tab:nxn} shows the result of this counting experiment. The most significant $n \times n$ bin is for $n = 10$; the probability for observing 29 or more events given 18.03 expected is 0.01. (This assumes systematic uncertainty on the background.) Thus we conclude that there is no statistically significant excess in the far tails of $\sht$ and $M_{reco}$.
\begin{table}[h]
  \begin{tabular}{c|c|c} \hline
    $m(\tprime)$ (GeV) & expected limit (pb) & observed limit (pb) \\ \hline
    180 & 2.954 $^{+0.818}_{-0.592}$ & 3.759 \\
    200 & 1.959 $^{+0.869}_{-0.525}$ & 1.595 \\
    220 & 0.693 $^{+0.309}_{-0.207}$ & 0.355 \\
    240 & 0.406 $^{+0.152}_{-0.100}$ & 0.258 \\
    260 & 0.288 $^{+0.102}_{-0.072}$ & 0.254 \\
    280 & 0.208 $^{+0.072}_{-0.047}$ & 0.237 \\
    300 & 0.150 $^{+0.052}_{-0.034}$ & 0.188 \\
    320 & 0.112 $^{+0.044}_{-0.031}$ & 0.165 \\
    340 & 0.088 $^{+0.032}_{-0.024}$ & 0.133 \\
    360 & 0.070 $^{+0.026}_{-0.020}$ & 0.112 \\
    380 & 0.056 $^{+0.021}_{-0.016}$ & 0.109 \\
    400 & 0.045 $^{+0.017}_{-0.013}$ & 0.081 \\
    450 & 0.033 $^{+0.012}_{-0.008}$ & 0.083 \\
    500 & 0.028 $^{+0.011}_{-0.006}$ & 0.073 \\ \hline
  \end{tabular}
  \caption{Expected and obtained limits on $\tprime$ production cross section for given mass.}
  \label{tab:result}
\end{table}
\begin{table}[t]
\begin{tabular} {ccc|ccc} \hline
n & Min $M_{rec}$ & Min $\sht$ & observed & expected & p-value \\
  & [GeV/c$^{2}$] & [GeV] &  &  & \\ \hline
1 & 475 & 775 & 0 & 0.021 & 1.000 \\
2 & 450 & 750 & 0 & 0.116 & 1.000 \\
3 & 425 & 725 & 1 & 0.228 & 0.2040 \\
4 & 400 & 700 & 2 & 0.371 & 0.0540 \\
5 & 375 & 675 & 3 & 0.718 & 0.0364 \\
6 & 350 & 650 & 4 & 1.503 & 0.0660 \\
7 & 325 & 625 & 4 & 2.876 & 0.3251 \\
8 & 300 & 600 & 12 & 5.498 & 0.0110 \\
9 & 275 & 575 & 14 & 9.885 & 0.1273 \\
10 & 250 & 550 & 29 & 18.03 & 0.0105 \\
11 & 225 & 525 & 41 & 31.34 & 0.0555 \\
12 & 200 & 500 & 58 & 52.05 & 0.2219 \\
13 & 175 & 475 & 92 & 91.14 & 0.4779 \\
14 & 150 & 450 & 152 & 158.7 & 0.7141 \\
15 & 125 & 425 & 222 & 231.0 & 0.7318 \\ \hline
\end{tabular}
\caption{Number of observed events in the highest $n \times n$ bins of $\sht$ and $M_{reco}$, compared with the prediction from a zero-signal fit to the full spectrum. For each value of $n$ the table shows the p-value, the probability for observing at least what was actually observed or more, given the number expected. The minimum $\sht$ and $M_{reco}$ values in each trial are also shown.}
\label{tab:nxn}
\end{table}


\bigskip 

\end{document}